\documentclass[a4paper,11pt]{article}
\usepackage{pos}
\usepackage{subfigure}
\usepackage{graphicx}
\usepackage{titlesec}
\titlespacing*{\section}
{0pt}{0.75em}{0.75em}

\let\oldbibliography\thebibliography
\renewcommand{\thebibliography}[1]{\oldbibliography{#1}
\setlength{\itemsep}{0pt}}

\title{Review of Atmospheric Neutrino Results from Super-Kamiokande}

\author{Volodymyr Takhistov\footnote{Speaker.} \textmd{\textit{(for the Super-Kamiokande Collaboration)}} 
}

\affiliation{Department of Physics and Astronomy, University of California, Los Angeles \\ Los Angeles, California, 90095-1547, USA}
\affiliation{Kavli Institute for the Physics and Mathematics of the Universe (WPI), UTIAS \\The University of Tokyo, Kashiwa, Chiba 277-8583, Japan}
 
\emailAdd{volodymyr.takhistov@ipmu.jp} 

\abstract{While neutrino physics enters precision era, several important unknowns remain. Atmospheric neutrinos allow to simultaneously test key oscillation parameters, with Super-Kamiokande experiment playing a central role. We discuss results from atmospheric neutrino oscillation analysis of the full dataset from Super-Kamiokande I-IV phases.  Further, we discuss tests of non-standard neutrino interactions with atmospheric neutrinos in Super-Kamiokande.}

\FullConference{
  40th International Conference on High Energy physics - ICHEP2020\\
  July 28 - August 6, 2020\\
  Prague, Czech Republic (virtual meeting)
}

\begin{document}
\maketitle

\section{Introduction}

Since the discovery of neutrino oscillations \cite{Fukuda:1998mi,Ahmad:2002jz}, stemming from the discrepancy of neutrino mass and their weak-interaction (flavor) eigenstates, significant progress has been made by reactor, atmospheric, solar as well as long-baseline experiments in measuring the fundamental oscillation parameters. Standard oscillations follow the three-flavor neutrino mixing paradigm based on the Pontecorvo-Maki-Nakagawa-Sakata (PMNS) matrix \cite{Pontecorvo:1967fh,Maki:1962mu}. This paradigm, described by unitary 3x3 transformation matrix $U_{\rm PMNS}$ relating flavor to mass eigenstates, is characterized by three mixing angles ($\theta_{12}$, $\theta_{23}$, $\theta_{13}$), two mass splittings ($\Delta m_{12}^2$, $\Delta m_{32}^2$), and
one CP-violating phase $(\delta_{\rm CP})$.

Precise determination of neutrino mass and mixing parameters is the principal goal of present-day neutrino oscillation experiments and is essential input for theoretical models, such as those based on flavor symmetries \cite{Feruglio:2019ktm} and Grand Unification \cite{Buchmuller:2002xm}.
While some of the oscillation parameters have been firmly established, several key unknowns remain: the ordering of the mass states with the largest splitting (hierarchy, characterized by sign of $\Delta m_{32}^2$), if the value of $\theta_{23}$ is slightly larger or smaller than $\pi/4$ ($\theta_{23}$ octant), the value of $\delta_{\rm CP}$.  Possibility of CP-violation in the lepton sector has attracted particular interest, due to potential association with the observed cosmological matter - anti-matter asymmetry, especially in models based on leptogenesis mechanism \cite{Fukugita:1986hr}. 

Atmospheric neutrinos provide a central arena for exploration of unknown oscillation parameters due to presence of both neutrinos and anti-neutrinos, oscillation matter effects as well as wide variety of energies and propagation pathlengths that can be probed. Super-Kamiokande (Super-K, SK) large water Cherenkov experiment is a leading experiment for studying atmospheric neutrinos.  

\section{Super-Kamiokande Experiment}

Super-Kamiokande is a 50 kiloton water Cherenkov detector (22.5 kiloton fiducial volume) located beneath a one-km rock overburden (2.7 km water-equivalent) within the Kamioka mine in Japan.
The detector is composed of an inner (ID, 11,146 inward-facing 20-inch PMTs, providing 40\% photo-coverage) and an outer (OD, 1,855 8-inch outward-facing PMTs) detector, which are optically separated. Cherenkov radiation, produced by charged particles traversing water, is
collected by the PMTs and is used to reconstruct physics events. The OD primarily serves as veto.

SK experiment has collected data during the SK-I (May 1996-Jul. 2001, 1489.2 live days), SK-II\footnote{An accident resulted in SK-II photo-coverage being reduced to 20\%. Full photo-coverage was restored in SK-III.} (Jan. 2003- Oct. 2005, 798.6 live days), SK-III (Jan. 2003- Oct. 2005, 518.1 live days) and SK-IV\footnote{Electronics have been upgraded in SK-IV.} (Sep. 2008- May 2018, 3244.4 live days) phases, with the latter ending in order to upgrade the detector. This corresponds to a combined exposure of 372.6 kton$\cdot$years. The details of detector design, performance, calibration, data reduction and simulations can be found in Ref. \cite{Fukuda:2002uc,Abe:2013gga}.

Super-Kamiokande is a highly versatile multi-purpose experiment, with capability to explore a broad variety of topics in the MeV - TeV energy range. This includes, among others, physics related to solar and atmospheric neutrinos, supernovae neutrinos, diffuse supernovae neutrino background, dark matter, proton decay and other baryon number-violating processes, as well as neutrino astrophysics.

\begin{figure*}[t]
\begin{center}
\includegraphics[trim={0mm 0mm 0 0},clip,width=.29\textwidth]{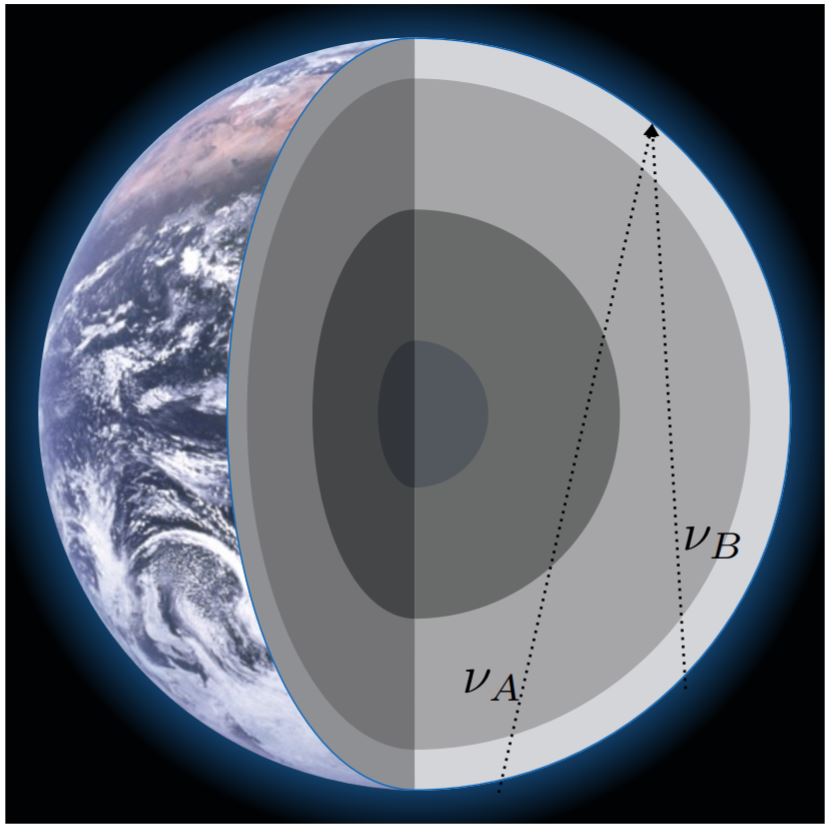}
\includegraphics[trim={0mm 0mm 0 0mm},clip,width=.34\textwidth]{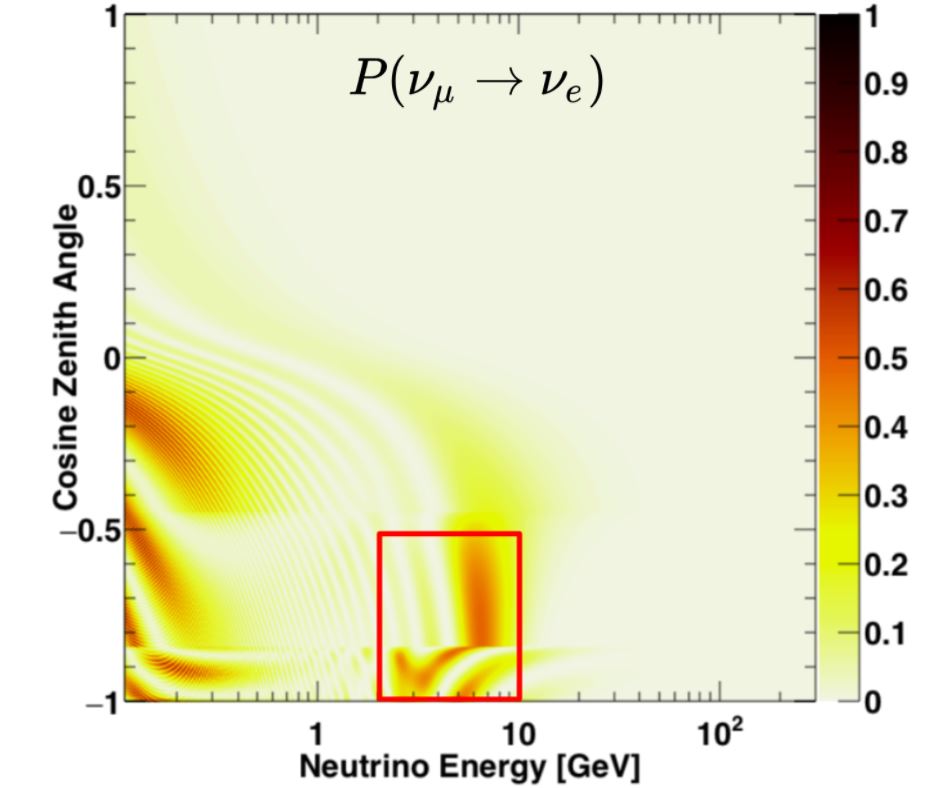}
\includegraphics[trim={0mm 0mm 0 0mm},clip,width=.33\textwidth]{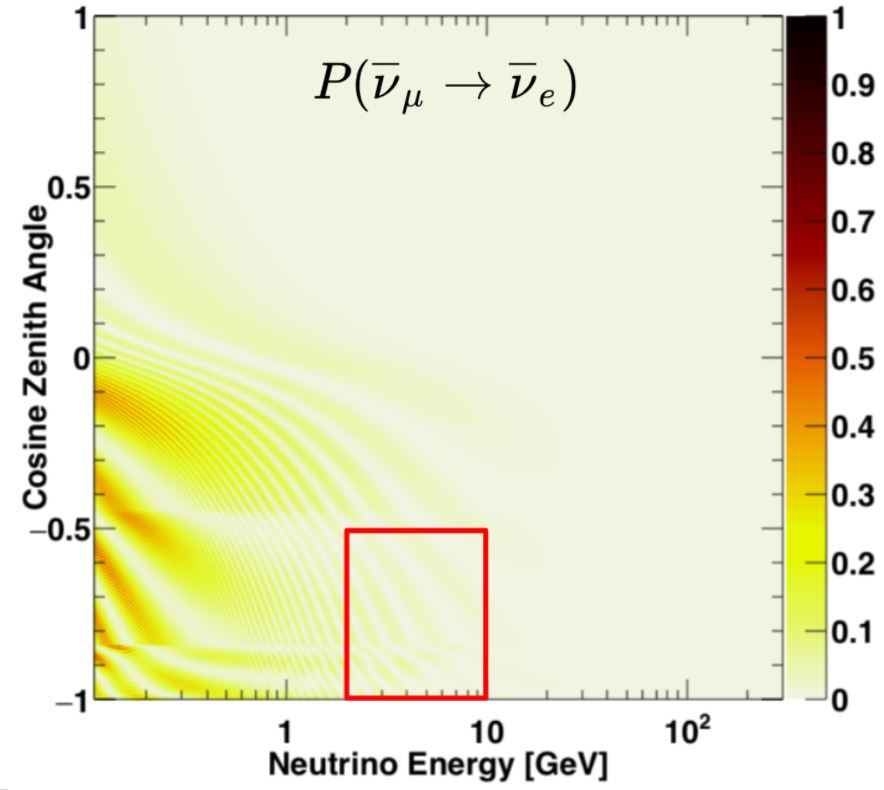}
\caption{ \label{fig:numat} 
\textbf{[Left]} Propagation of atmospheric neutrinos $\nu_A$ and $\nu_B$ produced from cosmic ray collisions with the atmosphere through varying matter density profile of Earth. \textbf{[Middle, Right]} Matter effects (red box) seen in up-ward going atmospheric $\nu_{\mu}$ and $\overline{\nu}_{\mu}$. Reproduced from Ref. \cite{Abe:2017aap}. 
}
\end{center}
\end{figure*}

\section{Standard Atmospheric Neutrino Oscillations}

Isotropic cosmic ray flux (primarily protons) hitting the atmosphere provides a permanent isotropic flux of $\nu_{\mu}$, $\overline{\nu}_{\mu}$, $\nu_e$, $\overline{\nu}_e$ atmospheric neutrinos with energies $E_{\nu} \sim 0.1 - 10^{2}$ GeV from meson decays in showers. Neutrino oscillations depend on energy and distance. Collecting atmospheric neutrinos from all directions, Super-K can probe neutrinos of $\sim 10 - 10^4$ km propagation baseline. Atmospheric neutrino interactions in the SK detector are simulated using the flux calculations of Honda \textit{et al.} \cite{Honda:2011} and the NEUT \cite{Hayato:2002sd} neutrino interaction software. Detector simulations based on GEANT-3 track particles resulting from the interactions \cite{Brun:1994aa}.

Oscillation parameters\footnote{ CP-symmetry is tested by comparing oscillation probability between neutrinos $P(\nu_{\alpha} \rightarrow \nu_{\beta})$ and anti-neutrinos  $P(\overline{\nu}_{\alpha} \rightarrow \overline{\nu}_{\beta})$.} can be studied via appearance or disappearance of particular neutrino species. SK discovered  neutrino oscillations by observing deficit of atmospheric upward-going vs. downward-going $\nu_{\mu}$'s \cite{Fukuda:1998mi}, implying oscillation into $\nu_\tau$'s. Subsequently, SK has detected \cite{Abe:2012jj} the appearance of $\nu_\tau$'s and measured their charged-current cross-section \cite{Li:2017dbe}. 

Neutrinos propagating through Earth with varying matter densities\footnote{Density profile accounted for via a simplified version of the preliminary
reference Earth model (PREM), see \cite{Abe:2017aap}.} (see Fig.~\ref{fig:numat}) experience resonant-like oscillation matter effects, due to modification of effective Hamiltonian stemming from difference in forward scattering of $\nu_e$ and $\nu_{\mu, \tau}$. This is characterized via $H_{\rm eff} = U_{\rm PMNS} M U_{\rm PMNS}^{\dagger} + V_e$, where $M$ is the neutrino mass-matrix and the additional potential term is $V_e = \sqrt2G_FN_e $, with $G_F$ being the Fermi constant and $N_e$ the electron number density. For anti-neutrinos, $V_e$ changes sign. Matter resonance effects occur only for neutrinos if the hierarchy is normal (NH,  $\Delta m_{32}^2 > 0$) and only anti-neutrinos if it is inverted (IH, $\Delta m_{32}^2 < 0$). The effects, depicted on oscillograms of Fig.~\ref{fig:numat}, are most pronounced in the upward-going multi-GeV $\nu_e$'s and $\overline{\nu}_e$'s. Such events are associated with suppressed neutrino flux and complicated multi-particle final states, resulting in limited statistics and mis-reconstruction.

The SK atmospheric data is divided into fully contained (FC), partially contained (PC) and upward-going muon (UP$\mu$) events\footnote{FC events have vertex reconstructed within fiducial volume but little OD activity, PC events have OD activity (corresponding to exiting particles) and UP$\mu$ events are from muon neutrinos interacting within rock below detector and depositing energy in both OD and ID.}. The samples are further separated according to various characteristics, such as number of decay electrons, number of Cherenkov rings, energetics, if rings are showering (e-like) or non-showering ($\mu$-like). In SK-IV, neutron tagging ($\sim 25\%$ efficiency) has become feasible, allowing for statistical separation into $\nu/\overline{\nu}$ for single-ring events. In total, for oscillation analysis 78 data samples are used and over 150 systematic errors are accounted for, describing uncertainties on neutrino flux and cross-section models as well as detector performance. 

\begin{figure*}[t]
\begin{center}
\includegraphics[trim={0mm 0mm 0 0},clip,width=.33\textwidth]{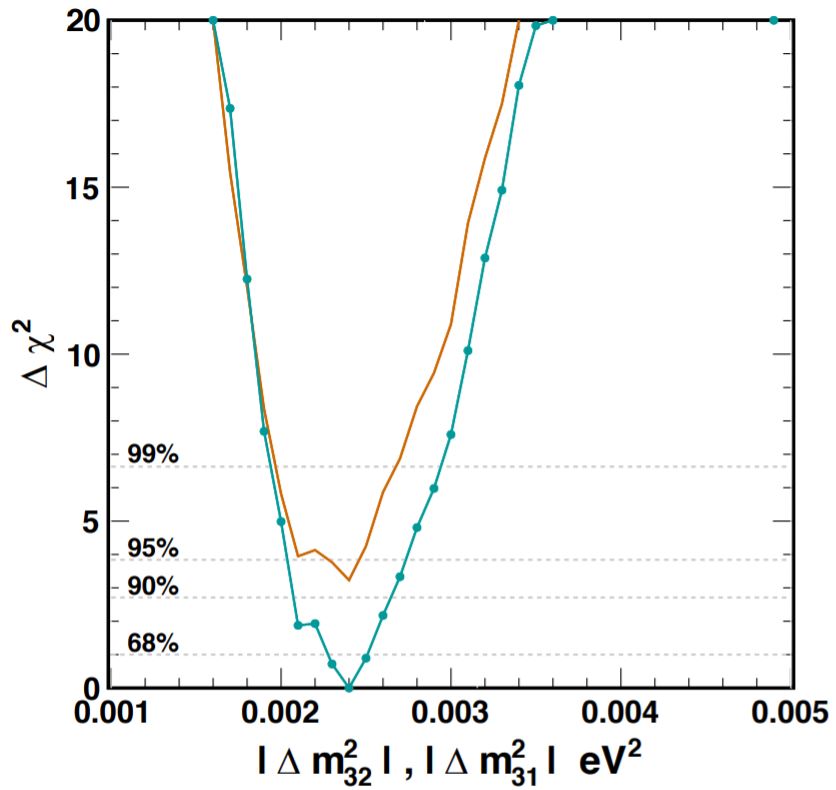}
\includegraphics[trim={0mm 0mm 0 0},clip,width=.32\textwidth]{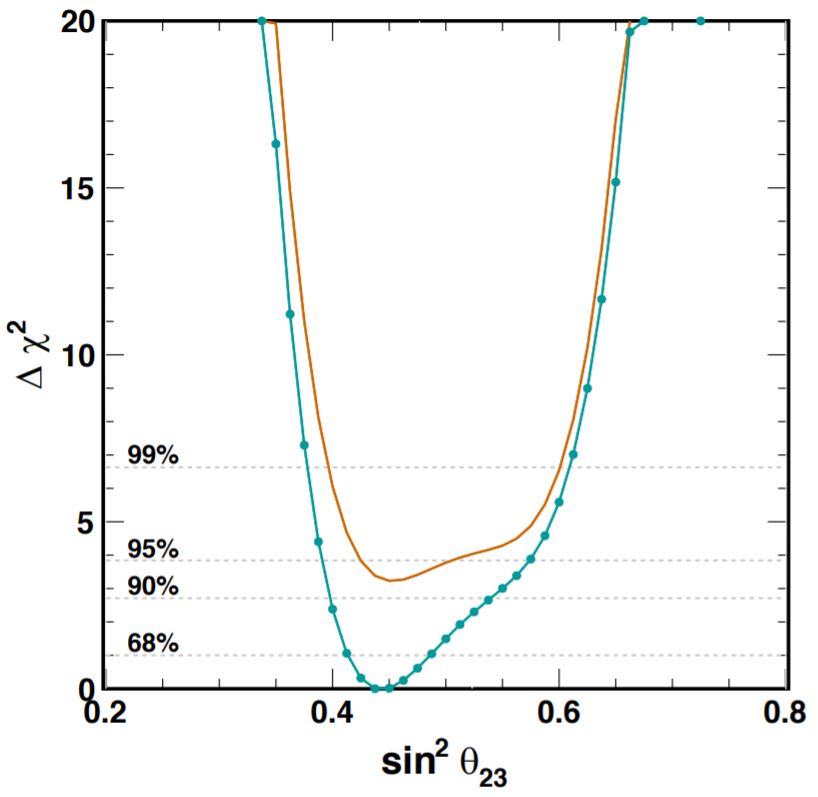}
\includegraphics[trim={0mm 0mm 0 0},clip,width=.32\textwidth]{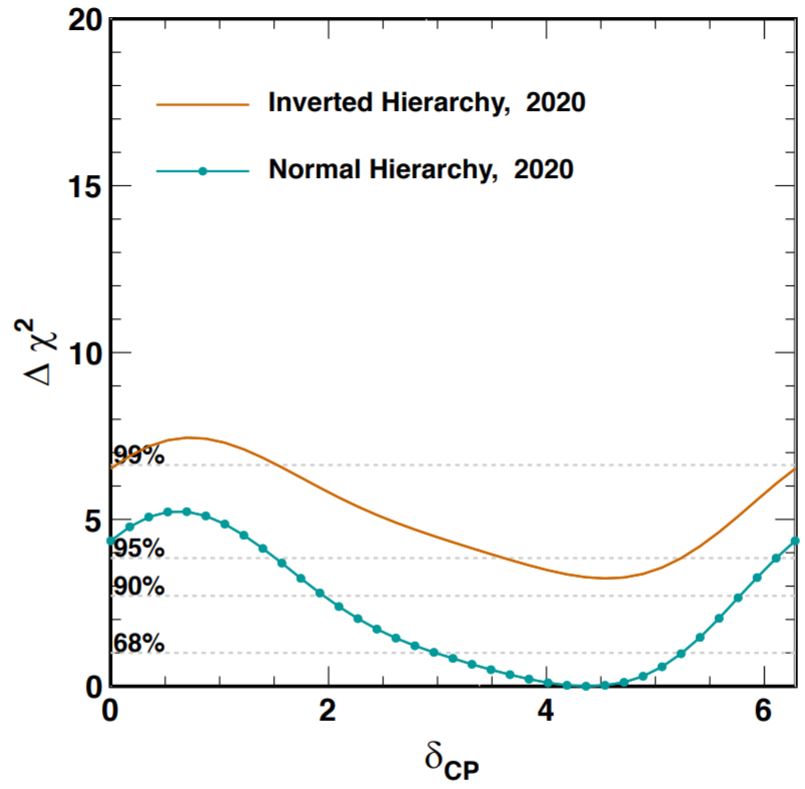}
\caption{ \label{fig:nufit} 
Atmospheric neutrino analysis fit results for full SK I-IV dataset, with fixed $\theta_{13}$. Inverted hierarchy result (orange) is offset from the normal hierarchy result (blue) by the difference in their minimum $\chi^2$ values.
}
\end{center}
\end{figure*}

Atmospheric oscillation analysis is performed with a simultaneous fit to $\theta_{23}$, $\Delta m_{23}^2$ and $\delta_{\rm CP}$ for each hierarchy assumption. The value of $\theta_{13}$ is constrained to $\sin^2
\theta_{13} = 0.0218$ as measured by reactor experiments \cite{Tanabashi:2018oca}, allowing for improved sensitivity (for an ``unconstrained analysis'' see Ref. \cite{Abe:2017aap}). The fit results for the full SK I-IV dataset are shown on Fig. \ref{fig:nufit} and Tab. \ref{tab:nubestfit}. The SK data disfavors IH at 71.4-90.3 \% C.L. and prefers 1st octant for $\theta_{23}$ as well as $\delta_{\rm CP} \sim 3 \pi/2$. These results improve upon the SK analysis of Ref. \cite{Abe:2017aap} with slightly less preference for NH. Enhanced constraints on the fit can be gained by including data from other experiments, such as the long-baseline T2K.

\begin{table*}[t]
  \centering
  \begin{tabular}{l|ccccc}
  \hline \hline
  Fit (930 bins) & $\chi_{\rm min}^2$ & $\theta_{13}$ & $\delta_{\rm cp}$ & $\theta_{23}$ & $\Delta m_{23} (\times  10^{-3})$\\
  \hline
    Normal Hierarchy & 1037.5 & 0.0218 & $4.36\substack{+0.88 \\ -1.39}$ & $0.44\substack{+0.05 \\ -0.02}$ & $2.40\substack{+0.11 \\ -0.12}$ \\
    Inverted Hierarchy & 1040.7 & 0.0218 & $4.54\substack{+0.88 \\ -1.32}$ & $0.45\substack{+0.09 \\ -0.03}$ & $2.40\substack{+0.09 \\ -0.32}$ \\ 
    \hline \hline
  \end{tabular}
  \caption{Summary of best fit results for atmospheric neutrino analysis of the full SK I-IV dataset, assuming either inverted or normal hierarchy. The p-value corresponding to such $\chi_{\rm min}^2$ is $\sim 0.32$.}
  \label{tab:nubestfit}
\end{table*}

\section{Exotic Scenarios}
 
With more precise neutrino oscillation parameter measurements, there is increased sensitivity to possible sub-dominant effects and new physics. One theoretically motivated class constitutes non-standard neutrino interactions (NSI, see \cite{Dev:2019anc} for review),  parametrized via non-renormalizable and non-gauge-invariant effective four fermion operators $\mathcal L_{\rm NSI} =-2\sqrt2G_F \epsilon_{\alpha\beta}^{fP}(\bar\nu_\alpha\gamma^\mu P_L\nu_\beta)(\bar f\gamma_\mu Pf')$, where $P$ are chiral projectors. NSI naturally appear in models with new light mediators and heavy states, however there are stringent constraints from lepton-flavor violation and universality. Such interactions will modify the effective Hamiltonian as
\begin{equation}
H_{\rm eff}=
U_{\rm PMNS}
\begin{pmatrix}
0\\&\frac{\Delta m^2_{21}}{2 E}\\&&\frac{\Delta m^2_{31}}{2 E}
\end{pmatrix}U_{\rm PMNS}^\dagger+V_e
\begin{pmatrix}
1+\epsilon_{ee}&\epsilon_{e\mu}&\epsilon_{e\tau}\\
\epsilon_{e\mu}^*&\epsilon_{\mu\mu}&\epsilon_{\mu\tau}\\
\epsilon_{e\tau}^*&\epsilon_{\mu\tau}^*&\epsilon_{\tau\tau}
\end{pmatrix}
\label{eq:nsi hamiltonian}
\end{equation}
where 1 in the $1+\epsilon_{ee}$ term is due to the standard charged current matter potential. Hence, NSI can alter propagation in matter. 

\begin{figure*}[t]
\begin{center}
\subfigure{
        \centering
        \includegraphics[width=.45\textwidth]{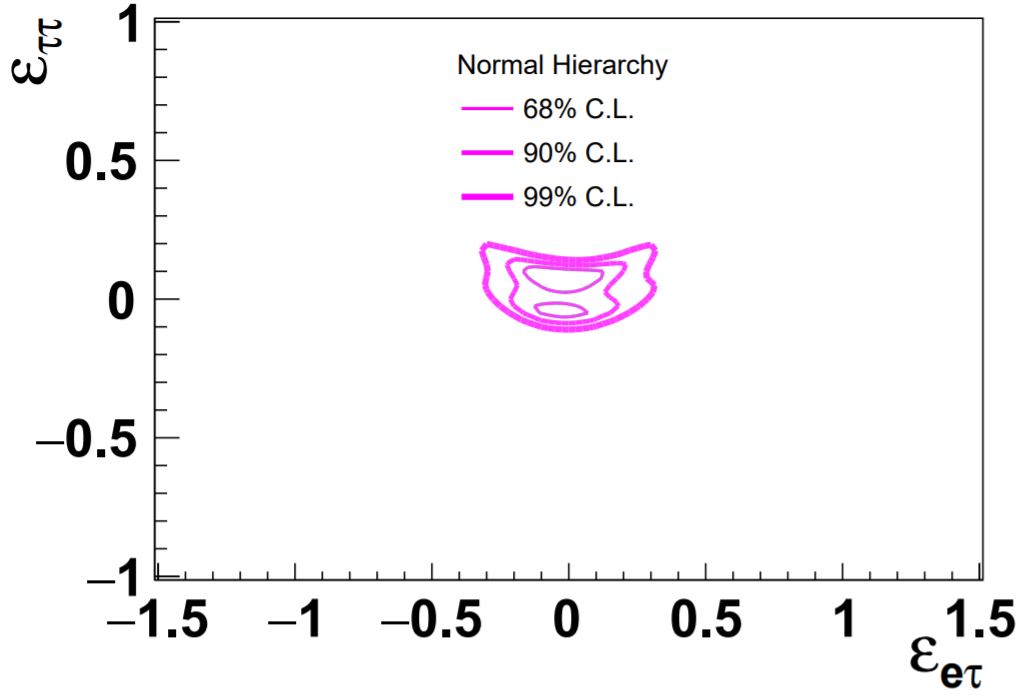}} 
\hspace{1cm}
\subfigure{
        \centering
        \includegraphics[width=.45\textwidth]{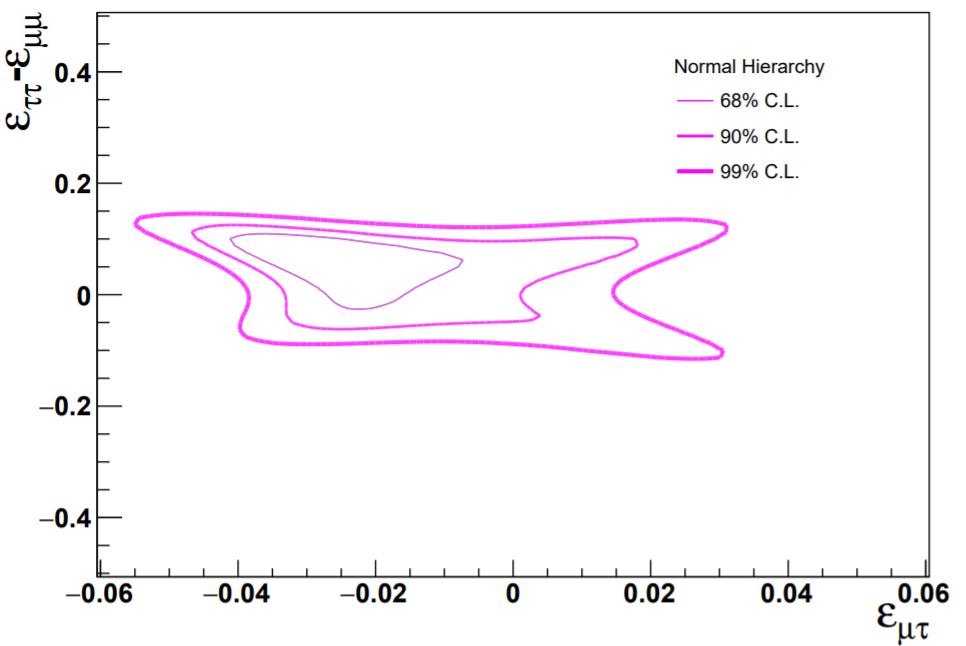}} 
\caption{ \label{fig:nsi} 
\textbf{[Left]} Best fit NSI in $e-\tau$ sector, assuming $\epsilon_{ee} = 0$ and NH.
\textbf{[Right]} Best fit NSI in $\mu-\tau$ sector, assuming NH.
}
\end{center}
\end{figure*}

SK can look for modifications in $\mu-\tau$ sector (affecting $\nu_{\mu}$ disappearance) and $e-\tau$ sector (affecting $\nu_{\mu}$ disappearance and $\nu_e$ appearance). To test $\epsilon_{\alpha\beta}$, standard oscillation parameters are fixed to best fit values. Fit results, which improve upon the previous SK I-II study \cite{Mitsuka:2011ty}, are depicted in Fig. \ref{fig:nsi} assuming NH\footnote{For IH, $\epsilon_{\alpha\beta} \rightarrow -\epsilon_{\alpha\beta}$.}. In the $\mu-\tau$ sector (with $\epsilon_{e \alpha}$ fixed), slight excess is observed in upward-going $\nu_{\mu}$'s above $\sim 100$ GeV ($\epsilon_{\mu\tau})$ and deficit in downward-going $\nu_{\mu}$'s in 20-80 GeV ($\epsilon_{\tau\tau})$. In the $e-\tau$ sector, slight excess is observed in upward-going $\nu_e$'s above $\sim 10$ GeV ($\epsilon_{e\tau})$ and deficit in downward-going $\nu_{\mu}$'s in 20-80 GeV ($\epsilon_{\tau\tau}$). The effects of $\epsilon_{\tau\tau}$ are similar to $\theta_{13}$, and with $\epsilon_{\mu\alpha} = 0$ follow relation $\epsilon_{\tau\tau} = |\epsilon_{e\tau}|^2/(1+\epsilon_{ee})$. The SK data are consistent with no contribution from NSI.

Employing atmospheric neutrinos SK can also probe other effects that modify oscillations, such as tests of the fundamental Lorentz invariance \cite{Abe:2014wla} or additional (sterile) neutrino states \cite{Abe:2014gda}.

\section{SK with Gadolinium (SK-GD)}

Gadolinium has been recently added to SK, starting SK-GD phase. At 0.2\% doping, this allows to improve neutron tagging efficiency up to 90\%. With the goal of first diffuse supernovae neutrino background detection \cite{Beacom:2003nk}, SK-GD boasts improvements for $\nu/\overline{\nu}$ and charged/neutral-current event separation as well as neutrino-energy reconstruction relevant for atmospheric neutrino studies. Further, SK-GD allows for improved sensitivity in studies where atmospheric neutrinos serve as background, such as for proton decay or indirect dark matter searches.

\section{Summary}

While neutrino oscillation studies are entering precision era, open questions remain. Atmospheric neutrinos allow to simultaneously test key oscillation parameters, with Super-Kamiokande playing a central role. Increased sensitivity is expected with the recently initiated SK-GD phase of experiment, and even further improvements from upcoming experiments like Hyper-Kamiokande.

\section*{Acknowledgement}
The author would like to thank the organizers for
the opportunity to present these results.  The Super-Kamiokande collaboration gratefully acknowledge the cooperation of the Kamioka Mining and Smelting Company. The Super-Kamiokande experiment has been built and operated from funding by the Japanese Ministry of Education, Culture, Sports, Science and Technology, the U.S. Department of Energy, and the U.S. National Science Foundation.

\bibliography{atmsk}

\bibliographystyle{JHEP.bst}

\end{document}